# On Teaching Quantum Physics at High School

## *By Enzo Bonacci[*]*

In the Italian education system, secondary students (ages 14-19) are confronted with the foundations of quantum physics during the final term of scientific high school (pre-university year). The Italian Ministry of Education, University and Research (acronym MIUR) has remarked its importance in the syllabus to address the high school exit examination (30% of the $5^{th}$ year physics course) but, due to limited learning time and intrinsic difficulty, this branch of physics is neither assimilated nor appreciated as it should. We wish to illustrate six didactic suggestions focused on learning motivation, emerged during a 17-year long teaching experience, which could help to tackle the main problems found. The key references are two peer-reviewed talks given, respectively, in 2018 at the 6th Annual International Conference on Physics by the Athens Institute for Education and Research and in 2013 at the $2^{nd}$ Rome workshop *Science Perception* by the Roma Tre University together with a concise and evocative poster outlining the history of quanta (Figure 1). Other useful resources are a 2015 conceptual diagram (Figure 2) and four invited lectures held in the years 2010-2017.

## Introduction

The twentieth century saw the affirmation of a physical theory nicely branded by the well-known Feynman's quote "I think I can safely say that nobody understands Quantum Mechanics" (Hey & Walters, 2003). Emblem of the Physics power to influence Philosophy and responsible for a good half of the hypotheses pervading the best science fiction movies, Quantum Mechanics is the second pillar of modern physics next to Einstein's relativity theory. Of this latter we know his aversion to the ontologically probabilistic character of the rival theory, expressed in the memorable correspondence with Max Born (Born & Einstein, 1971). Here we offer six didactic proposals which can positively address the numerous problems encountered in teaching basic Quantum Physics at science high schools. They come from two peer-reviewed talks (Bonacci, 2013a, 2018) together with a poster meant to be at the same time accessible and attractive to secondary school learners (Figure 1), whose *incipit* is Max Planck's study of Black Body Radiation (1900) but whose epistemological roots date back to ancient Oriental cultural traditions (Bonacci, 2013b). In what follows we will use interchangeably the terms Quantum Mechanics (QM for brevity) and Quantum Physics (QP for short) meaning the same course for high school students which does not go beyond the "first quantization" (semi-classical treatment). The legal framework and an up-to-date literature (Par. 2) are retrieved from current Italian institutional and sectorial websites.

[*]Teacher, Scientific High School "G.B. Grassi" of Latina, Italy.

*Figure 1.* The 2013 poster on Quantum Physics by Enzo Bonacci

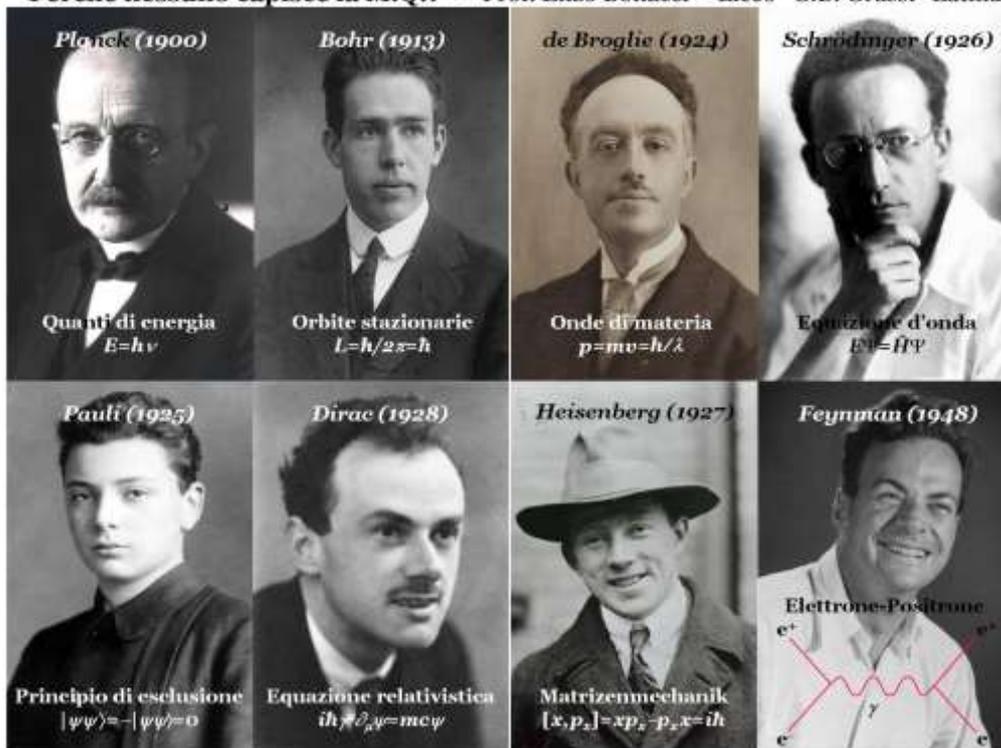

*Source*: https://bit.ly/2GBqq6J

## Physics in the Italian Scientific High School

### The Scientific High School in Italy

According to the EU's recent attention to Science, Technology, Engineering and Mathematics activities (acronym STEM), the Italian Ministry of Education, University and Research (MIUR) clarifies that "The course of the scientific high school promotes the acquisition of the knowledge and methods of mathematics, physics and natural sciences. It guides the student to deepen and develop knowledge and skills, to mature the skills necessary to follow the development of scientific and technological research and to identify the interactions among the different forms of knowledge, ensuring the mastery of the relative languages, techniques and methodologies, also through laboratory practice" (translated from the Italian website[1]).

---

[1] https://bit.ly/2GFr584.

**Italian National Guidelines for Physics at the Scientific High School**

The current study program of the "Liceo Scientifico" (Italian name for scientific high school) was established in the Decree of the President of the Italian Republic 89 of March 15, 2010 whose scheme of regulation was defined on October 7, 2010, by the Italian Interministerial decree n. 211. We extrapolate the text about the Physics course at the *Liceo Scientifico*: "At the end of the high school career the student will have learned the basic concepts of physics, the laws and theories that make them explicit, acquiring awareness of the cognitive value of the subject and the link between the development of physical knowledge and the historical and philosophical context in which it developed. In particular, the student will have acquired the following skills: observe and identify phenomena; formulate explanatory hypotheses using models, analogies and laws; formalize a physics problem and apply the mathematical and disciplinary tools relevant to its resolution; experience and explain the meaning of the various aspects of the experimental method, where the experiment is intended as a reasoned interrogation of natural phenomena, choice of significant variables, collection and critical analysis of data and reliability of a measurement process, construction and/or validation of models; understand and evaluate the scientific and technological choices concerning the society in which they live. The freedom, competence and sensitivity of the teacher - who will evaluate from time to time the most appropriate educational path for each class - will play a fundamental role in finding a connection with other teachings (in particular with those of mathematics, science, history and philosophy) and in promoting collaborations between his/her School and universities, research institutions, science museums and the world of work, mostly for the benefit of the students of the last two years" (translated from the Italian government's official journal[2]).

**Learning Objectives for Physics in the Last Year of the Liceo Scientifico**

The same Italian Interministerial decree n. 211 of October 7, 2010, elucidates the specific learning objectives of Physics in the *Liceo Scientifico*'s 5th year: "The student will complete the study of electromagnetism with magnetic induction and its applications, to arrive, favoring the conceptual aspects, to the synthesis formed by Maxwell's equations. The student will also study the electromagnetic waves, their production and propagation, their effects and their applications in the various frequency bands. The educational path will include the knowledge developed in the twentieth century related to the microcosm and the macrocosm, combining the problems that historically have led to new concepts of space and time, mass and energy. The teacher must pay attention to using a mathematical formalism accessible to students, always highlighting the founding concepts. The study of Einstein's theory of special relativity will lead the students to confront the simultaneity of events, the dilation of the times and the contraction of lengths; having faced the mass-energy equivalence will allow them to develop an energetic interpretation of nuclear phenomena (radioactivity, fission, fusion). The affirmation of the model of the quantum of light can be introduced through the study of thermal radiation and the Planck hypothesis (even just in a qualitative way), and will be developed on the one hand with the study of the photoelectric effect and its interpretation by Einstein, and on the other side with the discussion of theories and experimental results that highlight the presence of discrete energetic levels in the atom. Experimental evidence of the undulatory nature of matter, postulated by De Broglie, and the uncertainty principle could conclude the path significantly. The experimental dimension can be further explored with activities to be carried out not only in the educational laboratory of the school, but also in laboratories of universities and research institutions, also adhering to guidance projects. In this context, the students will be able to elaborate themes of their interest, approaching the most recent discoveries of physics (for example in the field of astrophysics and cosmology, or in the field of particle physics) or deepening the relationship between science and technology (for example the issue of nuclear energy, to acquire the scientific terms useful to critically approach the current debate, or semiconductors, to understand the most current technologies also in relation to the effects on the problem of energy resources, or micro- and nanotechnologies for the development of new materials)" (translated from the government's official journal[3]).

---

[2] https://bit.ly/2XOD7SH.

**Syllabus for Physics in the Last Year of the Liceo Scientifico**

The Physics course of the *Liceo Scientifico* is more detailed than in any other Italian secondary schools. We may summarize the syllabus of the 5th year as follows: "Electromagnetic induction. The Faraday-Neumann Law. The Law of Lenz. The currents of Foucault. Self-induction and mutual induction. The inductance of the solenoid. The RL circuit. Energy and energy density of the magnetic field. The alternating electromotive force and the alternating current. The ohmic circuit, the inductive circuit, the capacitive circuit. AC circuits. The LC circuit. The transformer. The transformation of tensions and currents. The electromotive force of a generator and the induced electromotive force. The induced electric field. Maxwell's equations. The electromagnetic field. Electromagnetic waves and flat electromagnetic waves. The relativity of space and time. Reference systems. The Michelson-Morley experiment. The axioms of the theory of special relativity. Simultaneity. The dilation of the times. The paradox of the twins. The beta coefficient and the gamma coefficient. The contraction of lengths. The Lorentz transformations. The Doppler effect of light. Redshift and blueshift. Special relativity. The space-time and the invariant interval between two events. The velocity addition. Mass-energy equivalence. The total energy of a relativistic particle. Kinetic energy, mass and relativistic momentum. Free fall and weightlessness. Acceleration and weight. The principles of general relativity. Non-Euclidean geometries. Geodetic curves. Gravity and the curvature of space-time. Gravitational deflection of light. The gravitational wave. From classical mechanics to quantum physics. Bohr's atom. Orbitals and quanta. Newton vs Huygens. Light both as particles and as waves: experimental evidences. The radiation of the black body. The photoelectric effect. The Heisenberg uncertainty principle. The superposition principle, the tunnel effect and the entanglement. Elements of nuclear physics and particle physics. Introduction to the Standard Model of Particles. Overview of astrophysics and cosmology" (translated from the Italian website[4]).

---

[3]https://bit.ly/2XOD7SH.

## Quantum Physics in the 5th Year of the Italian Scientific High School

The MIUR Decree n.10 of January 29, 2015, provided two important innovations in the *Liceo Scientifico*: the recognition of Physics, in addition to the traditional Mathematics, as "characterizing" the high school leaving certificate (http://www.gazzettaufficiale.it/eli/id/2015/02/24/15G00021/sg) and the integration of ministerial programs of the last year with modern Physics, in particular QM (Table 1).

*Table 1.* The QM Curriculum in the Italian Scientific Pre-University Year

| Category | Contents of the subject |
| --- | --- |
| Prerequisites | The Rutherford experiment and the atom model. Atomic spectra. Interference and diffraction (waves, optics). Discovery of the electron. Classic collisions. |
| Minimum essential topics | The emission of a black body and Planck's hypothesis. Lenard's experiment and Einstein's explanation of the photoelectric effect. The Compton effect. Bohr's model of atom and interpretation of atomic spectra. The Franck-Hertz experiment. Wavelength of De Broglie. Wave-particle duality; validity limits of the classical description. Diffraction / interference of electrons. The uncertainty principle. |
| Skills | Explain the model of the black body and interpret its emission curve based on the Planck model. Apply the laws of Stefan-Boltzmann and Wien. Apply the Einstein equation of the photoelectric effect for the resolution of exercises. Illustrate and know how to apply the Compton effect. Calculate the frequencies emitted by transition from Bohr's atom levels. Describe the quantization condition of the Bohr atom using the De Broglie relationship. Calculate the quantum indeterminacy on the position / momentum of a particle. Calculate the wavelength of a particle. Recognize the limits of classical treatment in simple problems. |
| Expertise | Knowing how to recognize the role of quantum physics in real situations and technological applications. |

*Source*: https://bit.ly/2GDOMNc.

---

[4]https://bit.ly/2LbqEqK.

## The Opinion of the Italian Pre-University Teachers of Physics

On November 4, 2015, MIUR in collaboration with the Dept. of Physics of the University "Roma Tre" published a survey conducted on a sample of 423 teachers (representative of the whole population) with the aim to test the reference frame of the Physics course of the last year at the *Liceo Scientifico*. We infer (Table 2) that the pre-university physics teachers deem more appropriate a program extended on electromagnetism (+5,73%) and reduced in both quantum physics (-3,7%) and advanced physics (-1,52%). Such response is symptomatic of that vast professional unease in teaching QP at secondary school level which motivated this paper together with the scarcity of pedagogical solutions. Even the *constructivism* fails with respect to the peculiarities of this branch of Physics (Karakostas & Hadzidaki, 2009). In fact, apart from marginal exceptions, the *learning by doing* is inapplicable.

*Table 2.* Testing the Pre-University Physics Curriculum Framework

| Module | Miur Guidelines | Teachers' Feedback |
| --- | --- | --- |
| Electromagnetism | 40% | 45,73% |
| Relativity | 20% | 19,49% |
| Quantum Physics | 30% | 26,3% |
| Advanced Physics | 10% | 8,48% |

*Source*: https://bit.ly/2Vo1LvH

## The 2018 Italian Reform of the Secondary School Leaving Exam

The MIUR Decree n. 769 of 26 November 2018 has established the "Reference frameworks for drafting and conducting written test" and the "evaluation grids for scoring" for the State Exams of the secondary school[5], enforcing the role of Physics in the Italian Scientific High School. Let us report characteristics and objectives of the exam in Physics at the *Liceo Scientifico*.

**Characteristics of the Physics exam at the Scientific High School.** The Physics test lasts 4-6 hours and consists in the solution of a problem chosen by the candidate between two proposals and in the answer to four questions among eight proposals (see the mock exam papers by MIUR[6]). "It is aimed at ascertaining the acquisition of the concepts and methods of physics with reference to the Fundamental Thematic Nuclei (Table 3) that vertically connect the topics covered in the course of study, in relation to the contents supplied by the current National Guidelines for the *Liceo Scientifico*. In particular, the test aims to detect the understanding and mastery of the scientific method and the capacity for physical argumentation through the use of hypotheses, analogies and physical laws. With reference to the various thematic cores, the solution of problems through the construction and discussion of models, the mathematical formalization, and the qualitative argumentation, the critical analysis of data may be requested in relation to natural phenomena or experiments. The test may contain references to classical texts or significant historical moments in physics" (translated from https://bit.ly/2RRt0xJ).

---

[5]https://bit.ly/2Vk7HpD
[6]https://bit.ly/2vul6gl.

*Table 3.* Fundamental Thematic Nuclei

| Modules | Units |
|---|---|
| Measurement and representation of physical quantities | Measurement uncertainty. Representations of physical quantities. |
| Space, Time and Motion | Kinematic quantities. Reference systems and transformations. Motion of a material point and a rigid body. Classical and relativistic kinematics. |
| Energy and Matter | Work and energy. Energy conservation. Energy transformation. Emission, absorption and transport of energy. |
| Waves and Particles | Sound and electromagnetic harmonic waves. Interference phenomena. Wave-particle dualism. |
| Forces and Fields | Representation of forces through the concept of field. Gravitational field. Electromagnetic field. Electromagnetic induction. |

*Source*: https://bit.ly/2PC4oVz.

**Objectives of the Physics exam at the Scientific High School.** With reference to the fundamental thematic nuclei (Table 3) and as expounded on the MIUR website[7], the Physics exam test aims to ascertain that the candidate is able to:

1. Represent, also graphically, the value of a physical quantity and its uncertainty in the appropriate units of measurement. Representing and interpreting, through a graph, the relationship between two physical quantities.
2. Evaluate the agreement between the experimental values of physical quantities in relation to measurement uncertainties in order to correctly describe the observed phenomenon.
3. Determine and discuss the motion of material points and rigid bodies under the action of forces.
4. Use the concept of center of mass in the study of the motion of two material points or of a rigid body.
5. Use the transformations of Galileo or Lorentz to express the values of kinematic and dynamic quantities in different reference systems.
6. Determine and discuss the relativistic motion of a material point under the action of a constant force or a Lorentz force.

---

[7]https://bit.ly/2FR3gfw.

7. Apply the relativistic relations on the dilation of the times and contraction of lengths and identify in which cases the non-relativistic limit is applied.
8. Determine the kinetic energy of a moving material point and the potential energy of a material point subjected to forces.
9. Link the variation of kinetic energy, potential energy and mechanical energy with the work done by the acting forces.
10. Use the conservation of energy in the study of the motion of material points and rigid bodies and in the transformations between work and heat
11. Use the conservation of energy in the study of the motion of material points and rigid bodies and in the transformations between work and heat.
12. Determine the energy density of electric and magnetic fields and apply the concept of energy transport by an electromagnetic wave.
13. Apply mass-energy equivalence in concrete situations taken from examples of radioactive decays, fission reactions or nuclear fusion.
14. Interpret the emission spectrum of the black body using the Planck distribution law.
15. Determine the frequencies emitted by transition between the energy levels of the Bohr atom.
16. Determine the wavelength, the frequency, the period, the phase and the speed of a harmonic wave and the relationships between these quantities.
17. Discuss interference phenomena with reference to sound or electromagnetic harmonic waves emitted by two coherent sources.
18. Discuss, also quantitatively, the wave-corpuscle duality.
19. Describe the quantization condition of the Bohr atom using the De Broglie relation.
20. Apply the Einstein equation of the photoelectric effect.
21. Describe the action of electric and magnetic gravitational forces through the concept of field. Represent an electric or magnetic field using the lines of force.
22. Employ the Gauss theorem to determine the characteristics of electric fields generated by symmetrical distributions of charges and to discuss the behavior of electrical charges in metals.
23. Employ the Ampère theorem to determine the characteristics of a magnetic field generated by a current wire and an ideal solenoid.
24. Describe and interpret electromagnetic induction phenomena and derive induced electromotive currents and forces.
25. Determine the force acting on an infinite-length current-carrying wire in the presence of a magnetic field, the force between two parallel current-carrying wires of infinite length and the force acting on a branch of a circuit moving in a magnetic field for the induced current. Determine the moment of the magnetic forces acting on a current-carrying loop in the presence of a uniform magnetic field.

**Six Problems in Teaching Quantum Physics**

The experience of seventeen years of teaching in a scientific high school allows to identify six macrocritical aspects that make learning Quantum Physics difficult for adolescents. They can be listed as follows:

1. The complexity of the QP theoretical system.
2. The standard probabilistic interpretation of QP.
3. An often misleading treatment of QP major themes in science-fiction.
4. The abstruseness of the mathematical formalism employed.
5. The multiplicity of approaches and discoveries structuring QP.
6. The low level of fame and/or charisma of the *Copenhageners*, i.e., the members of the so-called Copenhagen School.

How could we turn this series of apparently insurmountable obstacles to our advantage? We have found a solution through educational strategies to seize the corresponding opportunities summarized in the Table 4.

*Table 4.* Turning Six Problems into Opportunities while Teaching QP

| Problem | Opportunity |
|---|---|
| Discouraging conceptual intricacy | Appealing epistemological richness |
| Difficult ontological indeterminacy | Interdisciplinary study of chance |
| Abuse of QP terminology in sci-fi | Curiosity about QP basic notions |
| Worrying mathematical formalism | Compact QP formulation |
| Non-linear development of QP | Unifying explanations of QP logic |
| Copenhageners' low popularity | Captivating narration of QP story |

*Source*: https://bit.ly/2UTgcZ6.

**Six Proposals to Improve High School Education**

As highlighted by several researchers (Kohl, 2012), the peculiar world view advocated by Quantum Mechanics seems to have great affinity with old philosophical-religious traditions of the Indian subcontinent. Placing emphasis on this aspect, with further emotionally evocative stimuli from popular physics literature (Capra, 1977), might be of interest to pupils with a strong propensity towards introspective reflections, who usually show an apathetic detachment from scientific rationalism. The standard interpretation of $|\Psi|^2$ as the probability of finding the particle in a given volume element $dxdydz$ at time $t$ raises two reflections. The first concerns the measure paradox and implies an investigation of the QM conventional ontologies approachable merely as quick hints during the high school period. The second one, instead, implies the *Probability Calculus* which is curricular for Mathematics but does not temporally coincide neither with Physics nor with Chemistry. Probability could therefore be taught in three different moments: as an anticipation when atomic and molecular orbitals are introduced in Chemistry, as an exhaustive study in Mathematics and as a reference when dealing with QM in Physics. This redundancy, allowed by the interdisciplinary nature of Chance, surely benefits all the subjects involved. In order to avoid emotional barriers with the students passionate about science fiction and not to give a demoralizing impression of unintelligibility, it does not seem appropriate to correct immediately the inconsistencies related to quantum physics themes. Vice versa, it would be better to start from the zest that films and entertainment TV series arouse in young people to establish a common language and activate emotional intelligence (Parker, et al., 2004; Petrides, Frederickson, & Furnham, 2004). Only when the pupils have sufficient knowledge and a certain amount of interest towards QM we can return on detecting the science fiction's limits, through funny exercises like "find the mistake!". The heavy formalism adopted by QM can be useful in two moments: when we explicate the matrices in Mathematics, giving an outline of the *Matrizenmechanik*'s non-commutative algebra, and in the general introduction to the discipline, profiting from the extreme compactness of the formulas (combined with a qualitative elucidation of the topics) to diminish the fear in those who are about to study it. In this direction we put on a poster (Figure 1) eight of the fundamental formulas of the history of QM (Bonacci, 2013b). The illusion of a rapid comprehension of the equations will vanish progressively but without traumas, because in the meantime the students will have become acquainted with the most sophisticated mathematical operators. The tumultuous production of ideas in a very short time occurred for QM could make unsuccessful a chronological sequence. Even a classification by authors may not be convenient, since the same scientists returned to some questions repeatedly. The best report should be quasi-chronological, with small alterations permitting to build a logically sequential path. That is the reason why we put the Pauli exclusion principle in terms of state vector $|\psi\rangle$ next to the Schrödinger wave equation (1926), although it was formulated in 1925, and we placed the matrix mechanics of Heisenberg (1927) after Dirac (1928), i.e., the last author of the wave method (Figure 1). A schematic flowchart (Figure 2) might as well help the pupils to understand the logic behind the QM non-linear advancement (Bonacci, 2015a). We can remedy the lack of histrionics of the *Copenhageners*, included Niels Bohr (Clegg, 2013), in various ways: either with a partially anecdotal narration of the 30 years that "shook Physics" (Gamow, 1985) and drawing on the discreet repertoire of jokes uttered by eminent quantum physicists like by Feynman, already quoted in the Introduction (Hey & Walters, 2003), both with the use of images that stimulate the students' imagination. With this purpose, the eight young scientists on the poster (Figure 1) are pictured in poses revealing their different personalities. Between the firm gaze of Pauli and the nice exuberance of Feynman there is a whole range of expressions in which each student can recognize the closest to him/herself. This means creating a sort of empathetic identification with one or more authors and arousing that impelling curiosity that drives the supporters to know every detail of their idols' life and production.

**Results**

**The Pre-University Class 2007-2008**

In order to improve the effectiveness of our teaching Quantum Physics, we decided to change didactic method in the 2007-2008 school year; the pilot class was the VA of the Scientific High School "G.B. Grassi" in Latina. By adopting the strategies described in the third paragraph and taking advantage of some educational material available on the Internet (Bonacci, 2008a), we tried to manage three of the six critical aspects mentioned in the second paragraph:

1. The probabilistic interpretation of quantum phenomenology.
2. The rigorous mathematical formulation of QP.
3. The low level of notoriety of the Copenhageners.

We obtained surprisingly good results with a genuine transport of learners to the QP topics and their domination of the basic mathematical tools. After years in which Electromagnetism and Relativity had been the only subjects chosen by the students for the leaving certificate, four out of the twenty exam papers were centered on Quantum Mechanics. Such "innovative talks of undoubted quality", as declared by the President of the assessment committee, were entitled: "Simultaneous realities and parallel universes", "Paradox of Schrödinger's catv", "Quantum consciousness", "Schrödinger equation".

**The Pre-university Class 2008-2009**

Thanks to the different attitude of the following year VA students, a class with an evident predisposition to philosophical reflections and existentialist meditations, we tried to solve the other three critical issues that had in the meantime emerged in the teaching of Quantum Mechanics:

1. The epistemological ramifications of quantum theory.
2. An approximate representation of some QM issues in successful films.
3. The manifold contributions behind a unitary discipline.

After adopting the strategies described before, with the help of new on-line documents (Bonacci, 2008b), we noticed a huge interest, even fervid, towards the historical-epistemological aspect of matter, in spite of some weakness in calculation (at least compared to the technically perfect experience of the previous year). To confirm this, even 7 of the 24 exam essays were on QM-related topics and they were judged "original works of cultural depth and noteworthy interdisciplinary value" by the whole appraisal commission. The selected titles were: "Anthropic principle: weak, strong, participatory, final", "Quantum entanglement", "Schrödinger equation in the Copenhagen interpretation and the Everett's multiverse", "Quantum paradoxes: EPR, retro-causality, déjà vu, Schrödinger's cat", "Ontological vs. gnoseological uncertainty", "Revision of the concept of movement at the quantum level", "Quantum decoherence and role of the observer".

**The Pre-university Classes in the Years 2009-2018**

In the following years the material on contemporary Physics prepared for the pre-university classes has increased considerably thanks to:

- two invited lectures in the International Year of Astronomy 2009 (Bonacci, 2010a; 2010b);
- a photoelectric effect lab kit available since 2011;
- the activities of UniSchooLabS (http://unischoolabs.eun.org/);
- a peer-reviewed talk and poster at the *Science Perception* (Bonacci, 2013a; 2013b);
- a textbook aligned with the US physics program for grades 9-12 (Walker, 2014);
- a concept map for Italian-speaking students (Bonacci, 2015a), here translated into English (Figure 2);
- an invited lecture at the closing day of the Academic Year by the Astronomical Pontine Association (Bonacci, 2015b);
- an invited lecture in the *Aristotelian Paths* by the Italian Philosophical Society – Section of Latina "Feronia" (Bonacci, 2017);
- a peer-reviewed talk in the VI PHY ATINER (Bonacci, 2018);
- the resource repository of the SCIENTIX community for science education in Europe (http://www.scientix.eu/).

Once resolved the question of the sources, we tackled the critical aspects (always in the maximum number of three) trying different combinations with respect to 2–4–6 of the pilot year 2007-2008 and 1–3–5 of 2008-2009, but with overall lower marks. It seems indicative of an *optimum* result achievable only when some kind of problems (instead of others) are faced together, though we cannot state it with certainty. In this regard, we are looking forward to any feedback from colleagues who wish to implement our educational advice.

*Figure 2.* The 2015 Flowchart on Quantum Physics (translated into English)

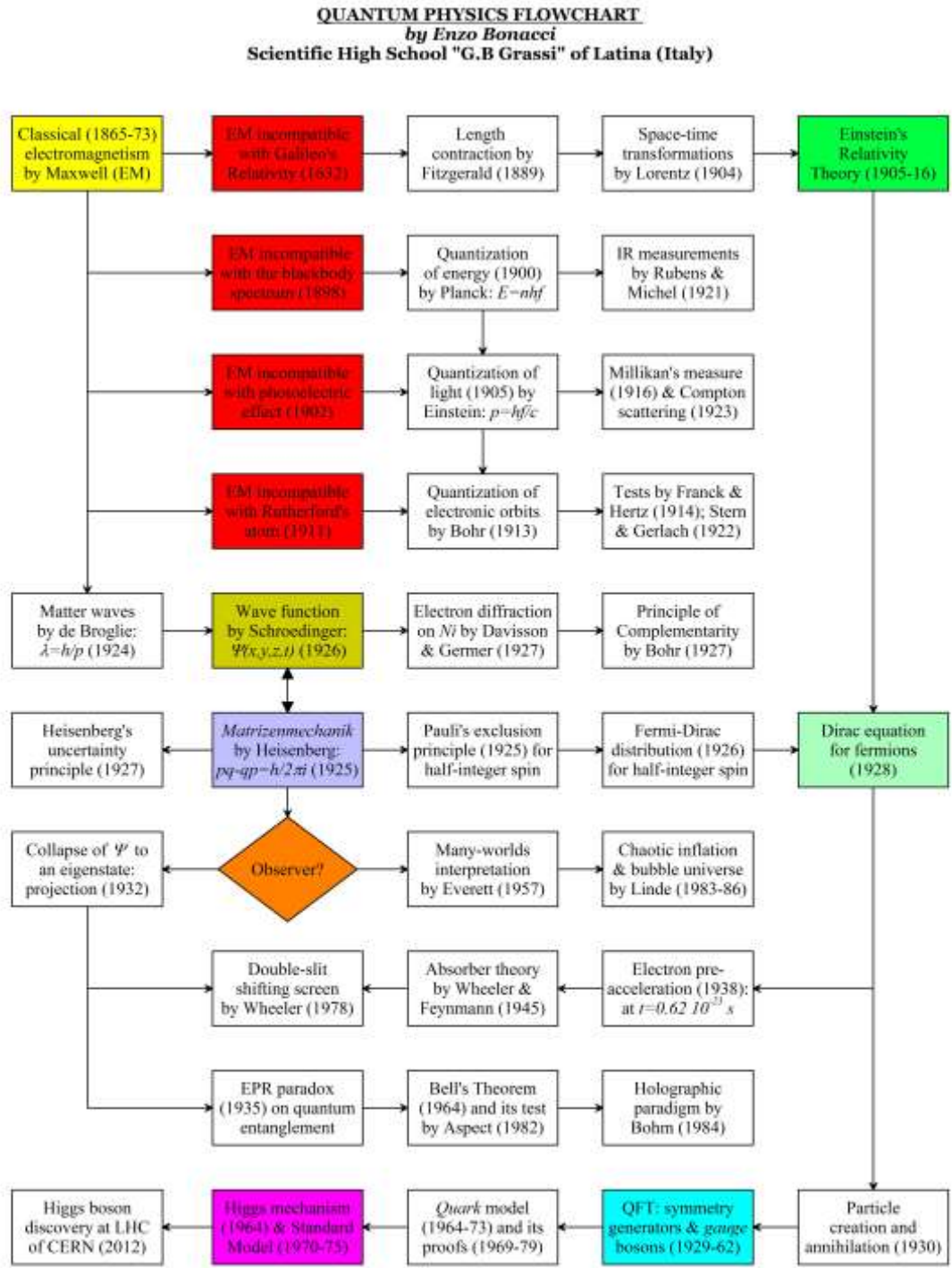

*Source*: https://bit.ly/2GG9tcw.

## Conclusions

In the tradition of the former project DESIRE (Disseminating Educational Science, Innovation and Research in Europe), which invited EU teachers to share their practical experiences and to integrate their teaching practices with inspiring tools (http://www.eun.org/projects/detail?articleId=706487), this paper has an empirical basis and is aimed to help concretely in an arduous task: teaching Quantum Physics at high school. It originates from the observation that introducing QP at secondary school can be such a tremendous shock that neither the finest educational psychology nor the most engaging textbook can avoid. The need for more empirical research into student difficulties and teaching strategies is underlined by the majority of literature focused on the improvement of the QP learning (Krijtenburg-Lewerissa, Pol, Brinkman, & van Joolingen, 2017). The educating community is called to supply *ad hoc* methods, efficacious to maximize the students' success. According to our working experience and personal re-elaboration, the high school teaching of Quantum Physics should benefit from:

1. A brief view on the Philosophy and Literature dealing with QP concepts;
2. A study of probability coordinated with Mathematics and Chemistry;
3. A gradual path of awareness about sci-fi: from enthusiasm to correction;
4. A familiarization with few fundamental QP formulas;
5. A scheme and a poster explaining the non-linear QP progress;
6. A popularizing narration of the QP vicissitudes and protagonists.

By virtue of the six didactic proposals advanced here, we should be able to overcome many of the obstacles we meet in our daily job. We should also be conscious of how utopian is to consider feasible the whole range of these professional tips, so that the selection of the most effective routes based on the needs and potential of the learners will ultimately be the true test of the pre-university teachers who will try themselves. We hope that an organic mosaic of experiential contributions on QP will enrich the current debate on STEM education (http://www.eun.org/it/focus-areas/stem) and, once consolidated this exploratory phase, such collective effort will eventually find its place in contemporary pedagogy.

## Acknowledgments

I'm grateful to the Organizing and Scientific Committee of the VI PHY 2018 for having given me the opportunity to share a hopefully interesting educational experience in the marvelous venue of the ATINER in Greece. I also thank the double-blind peer reviewers for their careful reading of the manuscript and their precious comments and suggestions.

**Appendix: Description of the 2013 Poster**

The 594x841 mm poster (Figure 1) consists of eight panels, arranged along two lines and four columns, each showing the face, in the foreground, of an important exponent of the QM with his surname, a contribution of him and the year of its introduction. While the educational reasons for this choice have already been widely ascertained, we are going to expound possible types of public presentation. Given that the ages of the photographed physicists do not necessarily correspond to the dates of publication of the formulas, whose succession is almost chronological to favor a thematic unification, the poster should be read from left to right and from top to bottom, that is, line by line. It covers the period 1900-1948, even if the main effort of defining the theory was accomplished in the first thirty years of the last century. In the first panel of the first line there is *Karl Ernst Ludwig Max Planck* and his equation on the energy quanta of 1900: $E = h\nu$. We may clear that it appeared, for the first time, in the black body radiation formula as the discrete energy of a single oscillator of the black cavity wall. In the second panel of the first line there is *Niels Henrik David Bohr* and his formula on the angular momentum of the electronic orbits allowed in an atom of 1913: $L = h/2\pi = \hbar$. One can clarify how the quantum condition for choosing stationary states is that the orbital angular momentum of the electron is an integer multiple of $\hbar$, a constant we will find in other formulas. In the third panel of the first row there is *Louis-Victor Pierre de Broglie* and his formula on the waves of matter of 1924: $p = mv = h/\lambda$. We can illustrate the analogy between the Undulatory Mechanics (based on the pilot waves of length $\lambda = h/p$) and the Undulatory Optics and then we may recall the first experimental confirmation in 1927 by Davisson and Germer with the electronic diffraction. In the fourth panel of the first row there is *Erwin Rudolf Josef Alexander Schrödinger* and his 1926 formula on the wave equation expressed in the form: $E\Psi = \hat{H}\Psi$. We can refer to the meaning of $E$ as the eigenvalue of the energy for the system, of $\hat{H}$ as the Hamiltonian operator for a harmonic quantum oscillator and of $\Psi$ as a wave function and we may clarify that the electronic population, corresponding to a certain level of energy (i.e., to a certain *eigenvalue* $E$) is represented by the eigenfunctions of the Hamiltonian operator, solutions of the equation. In the first panel of the second line there is *Wolfgang Ernst Pauli* and his 1925 exclusion principle in the form: $|\psi\psi\rangle = -|\psi\psi\rangle = 0$. We can mention how two half-integer spin particles (fermions) of the same species form totally antisymmetric states and the impossibility that they both occupy the same quantum state $|\psi\rangle$ because of the null ket. In the second panel of the second line there is *Paul Adrien Maurice Dirac* and his 1928 equation in the form: $i\hbar\gamma^\mu\partial_\mu\psi = mc\psi$. We could briefly say that it describes the motion of fermions in a relativistically invariant way, without further specification. We should however underline the theoretical prediction of the electron's antiparticle (positron), as well as the experimental confirmation of the positron obtained by Anderson in 1932 while analyzing the cosmic rays. In the third panel of the second line there is *Werner Karl Heisenberg* and the quantization condition of the *Matrizenmechanik* formulated in 1927: $[x, p_x] = xp_x - p_x x = i\hbar$. We may hint that the matrix mechanics describes the relation between the coordinate of position $x$ and the conjugated moment $p$ of a particle and that the indeterminacy $\Delta p \Delta x \geq \hbar/2$ descends from the quantization condition. In the fourth panel of the second row there is *Richard Phillips Feynman* with one of his homonymous diagrams (introduced in 1948) used to describe the annihilation and creation of the electron-positron pair: $e^+ + e^- \to \gamma \to e^+ + e^-$. One can point out the importance of Feynman diagrams in the description of any quantum interaction and the rapidity with which they were universally adopted. We may add another example $e^+ + e^- \to Z^0 \to \mu^+ + \mu^-$ allowing a connection to the two Italian Nobel Prizes Enrico Fermi (awarded in 1938) and Carlo Rubbia (awarded in 1984) as, respectively, starting and arrival point of the long path leading to the full comprehension of the electroweak interaction.